\documentstyle[nassau,pslatex,epsfig]{article}
\frompage{000} \topage{000}                                              
\def\rightmark{Two RHIC puzzles} 
\def\leftmark{U. Heinz and P.F. Kolb}
\def\eq{{\,=\,}}
\def\v{{\rm v}}

\title{Two RHIC puzzles: Early thermalization and the HBT problem}
\authors{Ulrich Heinz and Peter F. Kolb\\[2.812mm]
        {\normalsize
         Physics Department, The Ohio State University, Columbus,
         OH 43210, USA}}
 
\abstract{Hadron spectra from the first year RHIC run are shown to be
excellently reproduced by hydrodynamic calculations. We argue that in
particular the elliptic flow data provide strong evidence for early 
thermalization at RHIC, at energy densities well above deconfinement, 
but that the phenomenologically extracted short thermalization time 
scale of less than 1\,fm/$c$ provides a serious challenge for theory. 
The HBT radii from the hydrodynamic calculations agree only 
qualitatively with the data, showing significant quantitative 
discrepancies. It is argued that this points to a still incomplete 
understanding of the freeze-out process at RHIC.}


 
\begin{document}
 
\maketitle

\setcounter{page}{1}

\section{Elliptic flow as an early QGP signature}
\label{sec1}

The quark-gluon plasma (QGP) is a thermalized system and, as such, has 
thermal pressure. If the QGP is created in a heavy-ion collision, this 
pressure acts against the surrounding vacuum and causes a rapid 
collective expansion (``flow'') of the reaction zone, the ``Little Bang''. 
In non-central collisions the initial overlap region of the two nuclei
is elliptically deformed in the transverse plane, resulting in anisotropic
pressure gradients. These cause a more rapid expansion into the reaction
plane than perpendicular to it, resulting in an anisotropy of the
final transverse momentum distribution called {\em elliptic flow} 
\cite{O92}. Elliptic flow is quantified by the second harmonic coefficient 
$\v_2(y,p_\perp;b)$ of a Fourier expansion in $\phi_p$ of the measured 
hadron spectrum $dN/(dy\,p_\perp dp_\perp\,d\phi_p)$ \cite{VZ96}. At 
midrapidity in collisions between equal-mass nuclei it is the 
lowest-order non-zero Fourier coefficient.

Since individual nucleon-nucleon collisions produce azimuthally symmetric
spectra, any final state momentum anisotropies must be generated dynamically 
during the nuclear reaction. A non-zero value for the elliptic flow is thus 
an unambiguous sign for reinteractions among the produced particles, and its
magnitude characterizes their strength. Microscopic transport calculations 
\cite{Zhang:1999rs,Molnar:2001ux} show a monotonic dependence of $\v_2$ 
on the opacity (density times scattering cross section) of the produced 
matter which is inversely related to its thermalization time. Elliptic 
flow also requires the existence of an initial spatial anisotropy of 
the reaction zone, by colliding either deformed nuclei at $b\eq0$ or 
spherical nuclei at $b{\,\ne\,}0$. The transport calculations strongly 
suggest that, for a given initial spatial anisotropy $\epsilon_x$, the 
maximum momentum-space response $\v_2$ is obtained in the 
{\em hydrodynamic limit} which assumes a thermalization time which is 
much shorter than any macroscopic time scale in the system. Any 
significant delay of thermalization allows the initial spatial 
deformation to decay without concurrent build-up of momentum anisotropies, 
thereby reducing the finally observed elliptic flow signal
\cite{Kolb:2000sd}.

This specific sensitivity of the elliptic flow to rescattering and 
pressure build-up in the early collision stages 
\cite{Sorge:1997pc,Voloshin:2000gs} (before the spatial deformation and 
the resulting anisotropies of the pressure gradients have disappeared 
\cite{Kolb:2000sd}) puts $\v_2$ on the list of ``early signatures'' of 
the collision dynamics. In contrast to other early probes (which use 
rare signals such as hard photons and dileptons,
heavy quarkonia and jets), $\v_2$ can be extracted from the bulk of
the measured hadrons which are very abundant and thus easily accessible.
In fact, the elliptic flow measurement in Au+Au collisions at 
$\sqrt{s}\eq130\,A$\,GeV \cite{Ackermann:2001tr} became the {\em second} 
publication of RHIC data and appeared within days of the end of the first 
RHIC run.

We present hydrodynamic results for hadronic spectra and elliptic flow 
at RHIC energies. We show that the hydrodynamic approach provides an 
excellent quantitative description of the bulk of the data and fails 
only for very peripheral Au+Au collisions and/or at high 
$p_\perp{>}1.5{-}2$\,GeV/$c$. That the hydrodynamic approach fails for 
small initial overlap regions or large hadron transverse momenta is not 
unexpected. What is surprising is that the hydrodynamic approach {\em does} 
work for such a wide range of impact parameters and $p_\perp$ and 
quantitatively reproduces the momenta of more than 99\% of the particles: 
below $p_\perp\eq1.5$\,GeV/$c$ the elliptic flow data 
\cite{Ackermann:2001tr,Lacey:2001va,Poskanzer:2001cx} 
actually exhaust the hydrodynamically predicted 
\cite{Kolb:2000sd,Kolb:1999it,Teaney:2001cw,Kolb:2001fh,Huovinen:2001cy} 
upper limit. The significance of this agreement can hardly be overstressed, 
and it poses significant challenges for microscopic descriptions of the 
early collision dynamics. How the system is able to thermalize so fast
is the first RHIC puzzle which we point out. 

The second puzzle arises from two-particle correlation measurements 
(HBT interferometry) \cite{Adler:2001zd,Johnson:2001zi} which constrain 
the freeze-out distribution in space-time. The same hydrodynamic model 
which provides an almost perfect description of the momentum-space 
structure of the emitting source (spectra an elliptic flow) predicts 
a space-time distribution at freeze-out which, when translated into
HBT radii, does not agree very well with the data. We argue in 
Sec.~\ref{sec4} that this points to a problem with our understanding of 
the late freeze-out stage of the collision.

\section{Radial and elliptic flow from hydrodynamics}
\label{sec3}

The natural language for describing collective flow phenomena is 
hydrodynamics. In the ideal fluid (non-viscous) limit used by us, this 
approach assumes that the microscopic momentum distribution is thermal 
at every point in space and time. This does not require chemical 
equilibrium -- chemically non-equilibrated situations can be treated by 
solving separate and coupled conservation equations for the particle 
currents of individual particle species. The assumption of local thermal 
equilibrium is an external input, and hydrodynamics offers no insights 
about the equilibration mechanisms. It is clearly invalid during the initial 
particle production and early recattering stage, and it again breaks 
down towards the end when the matter has become so dilute that 
rescattering ceases and the hadrons ``freeze out''. The hydrodynamic 
approach thus requires a set of {\em initial conditions} for the 
hydrodynamic variables at the earliest time at which the assumption of
local thermal equilibrium is applicable, and a {\em ``freeze-out 
prescription''} at the end. For the latter we use the Cooper-Frye algorithm 
\cite{Cooper:1974mv} which implements an idealized sudden transition 
from perfect local thermal equilibrium to free-streaming. A better algorithm 
\cite{Bass:2000ib,Teaney:2001cw} switches from a hydrodynamic description 
to a microscopic hadron cascade at or shortly after the quark-hadron 
transition, before the matter becomes too dilute, and lets the cascade 
handle the freeze-out kinetics. The resulting flow
patterns \cite{Teaney:2001cw} from such an improved freeze-out algorithm 
don't differ much from our simpler Cooper-Frye based approach. 

The main advantage of the microscopic freeze-out algorithm 
\cite{Bass:2000ib,Teaney:2001cw} is that it also correctly reproduces 
the final chemical composition of the fireball, since the particle 
abundances already freeze out at hadronization, due to a lack of 
particle-number changing inelastic rescattering processes in the 
hadronic phase \cite{Heinz:1999kb}. Our version of the hydrodynamic 
approach uses an equation of state which assumes local chemical 
equilibrium all the way down to kinetic freeze-out at 
$T_{\rm f}{\,\approx\,}125$\,MeV and thus is unable to reproduce the 
correct hadron yield ratios. We therefore adjust the normalization of 
the momentum spectra for the rarer particle species (kaons, protons, 
antiprotons) in central collisions by hand to reproduce the chemical 
equilibrium ratios at a chemical freeze-out temperature 
$T_{\rm chem}\eq165$\,MeV. The absolute normalization of the pion 
spectra is adjusted through the initial energy density in central 
collisions; for non-central collisions no new parameters enter since 
the centrality dependence of the initial conditions is completely 
controlled by the collision geometry.

To simplify the numerical task of solving the hydrodynamic equations we 
analytically impose boost invariant longitudinal expansion 
\cite{Bjorken:1983qr,O92}. This doesn't give up any essential physics as 
long as we focus on the transverse ex\-pan\-sion dynamics near midrapidity 
(the region which most RHIC experiments cover best). The hydrodynamic 
expansion starts at time $\tau_{\rm eq}$ which we fixed by a fit to 
hadron spectra from the SPS and then extrapolated to RHIC initial 
conditions (for details see \cite{Kolb:2000sd,Kolb:2001fh}). For each 
impact parameter the initial energy density profile in the transverse plane
is calculated from a Glauber parametrization using realistic nuclear 
thickness functions \cite{Kolb:2000sd,KHHET}. The measured centrality
dependence of the charged particle rapidity density at midrapidity, 
$(dN_{\rm ch}/dy)(y\eq0)$, selects certain allowed combinations of 
``hard'' and ``soft'' mechanisms for the initial particle production 
\cite{KHHET}. We here present results for initial conditions at 
$\tau_{\rm eq}\eq0.6$\,fm/$c$ calculated from a mixture of 25\% 
``hard'' (binary collision) and 75\% ``soft'' (wounded nucleon) 
contributions \cite{KHHET} to the initial {\em entropy density} (or 
parton density), with a maximal entropy density $s_{\rm max}\eq85/$fm$^3$
at the fireball center in central collisions (corresponding to a maximal 
energy density $e_{\rm max}\eq21.4$\,GeV/fm$^3$ and a maximal temperature 
$T_{\rm max}\eq328$\,MeV). At the standard time $\tau\eq1$\,fm/$c$ used 
in Bjorken's formula \cite{Bjorken:1983qr} for estimating the energy 
density from the measured multiplicity density, this corresponds to an 
average energy density $\langle e\rangle(1\,{\rm fm}/c)\eq5.4$\,GeV/fm$^3$ 
which is about 70\% higher than the value reported from 158\,$A$\,GeV 
Pb+Pb collisions at the SPS. (Note that $\langle e\rangle$ at 
$\tau_{\rm eq}\eq0.6$\,fm/$c$ is nearly twice as large!) The 
corresponding profiles for peripheral collisions are then given by 
the Glauber model \cite{Kolb:2000sd,KHHET}. Kinetic freeze-out was 
forced at $T_{\rm f}\eq130$\,MeV, independent of centrality.

\begin{figure}[htb]
\epsfig{file=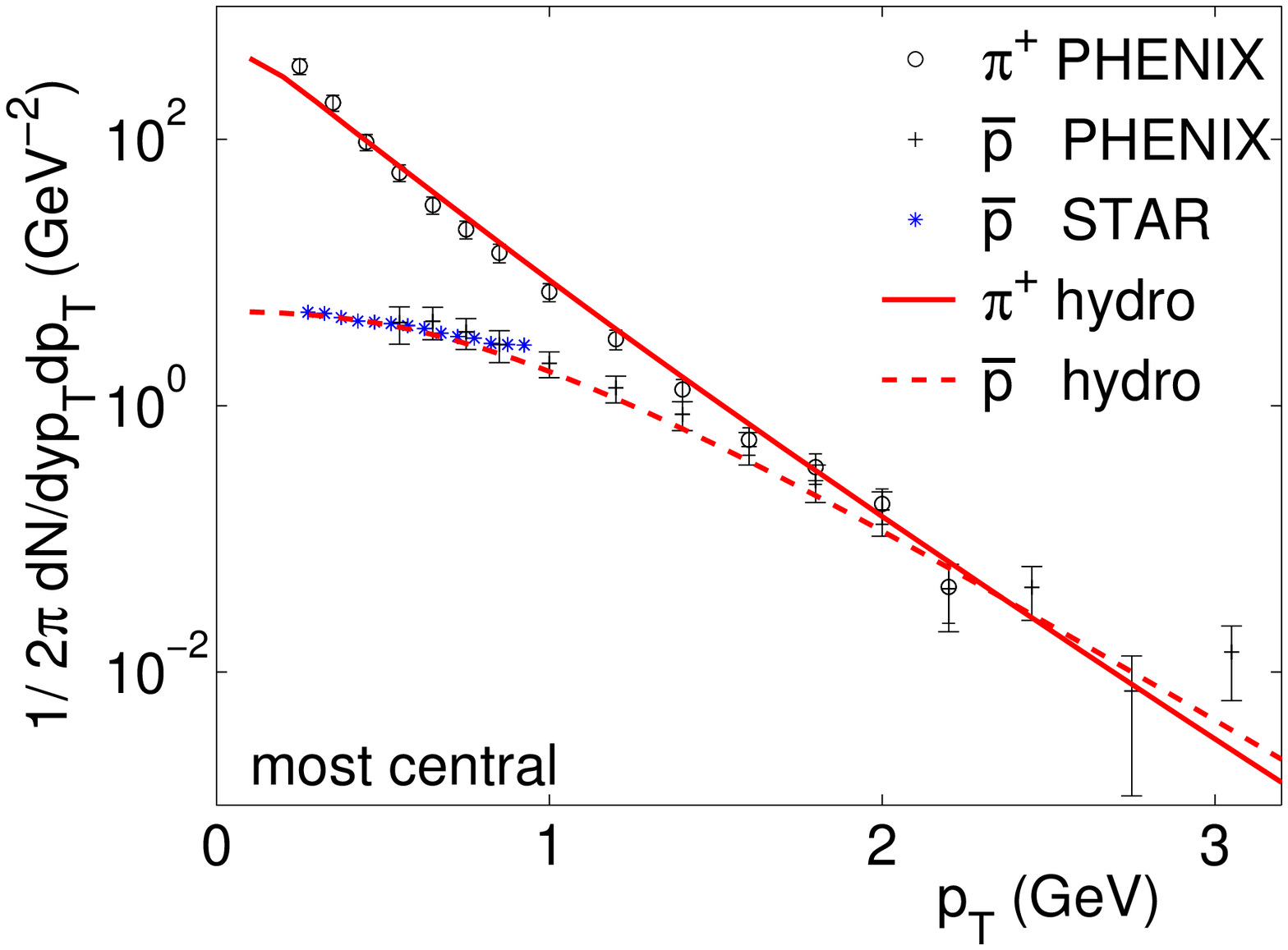,width=60mm,height=50mm}\hspace*{4.5mm}
\epsfig{file=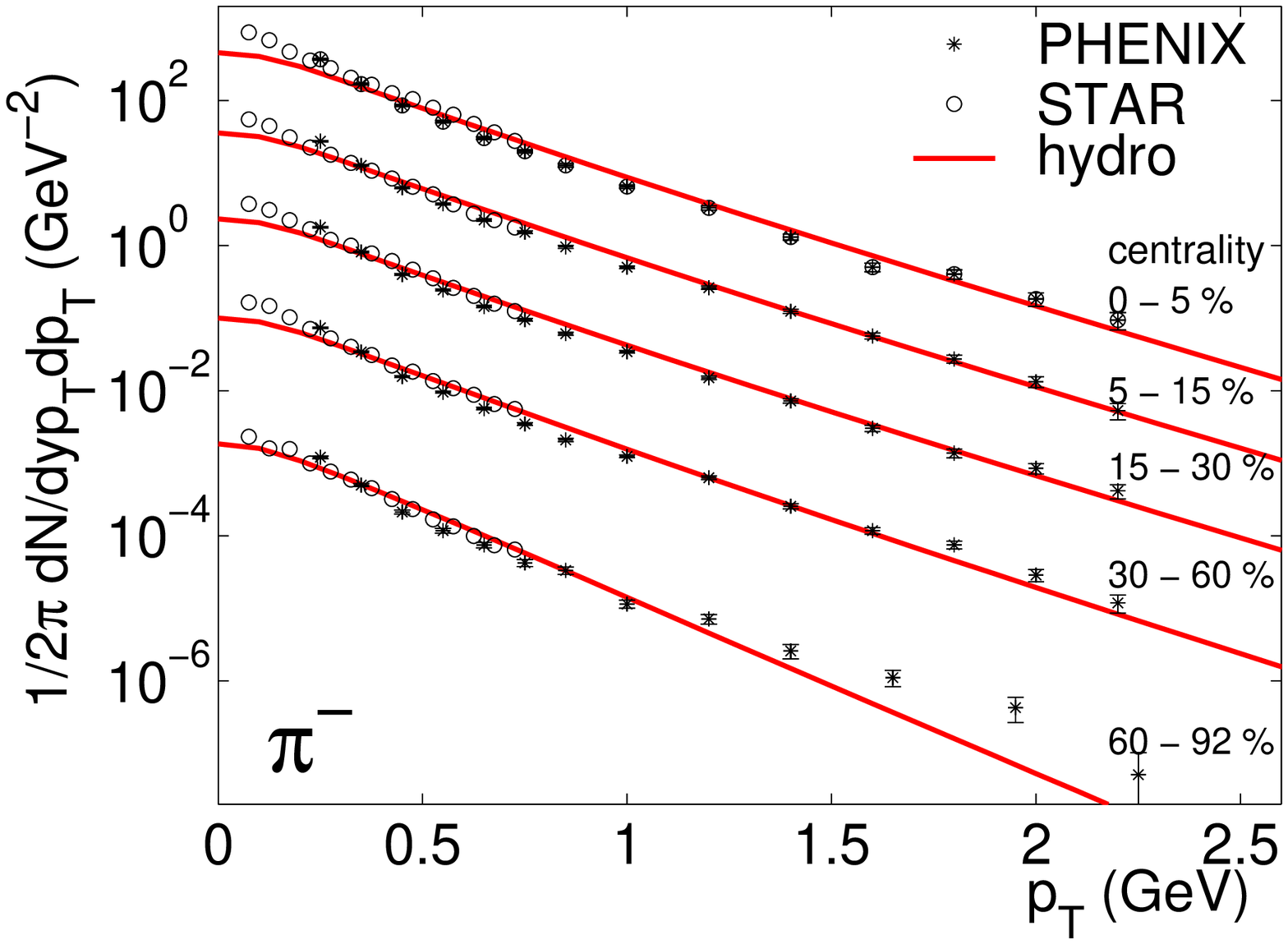,width=59mm,height=50mm}\hfill{\phantom{n}}
\epsfig{file=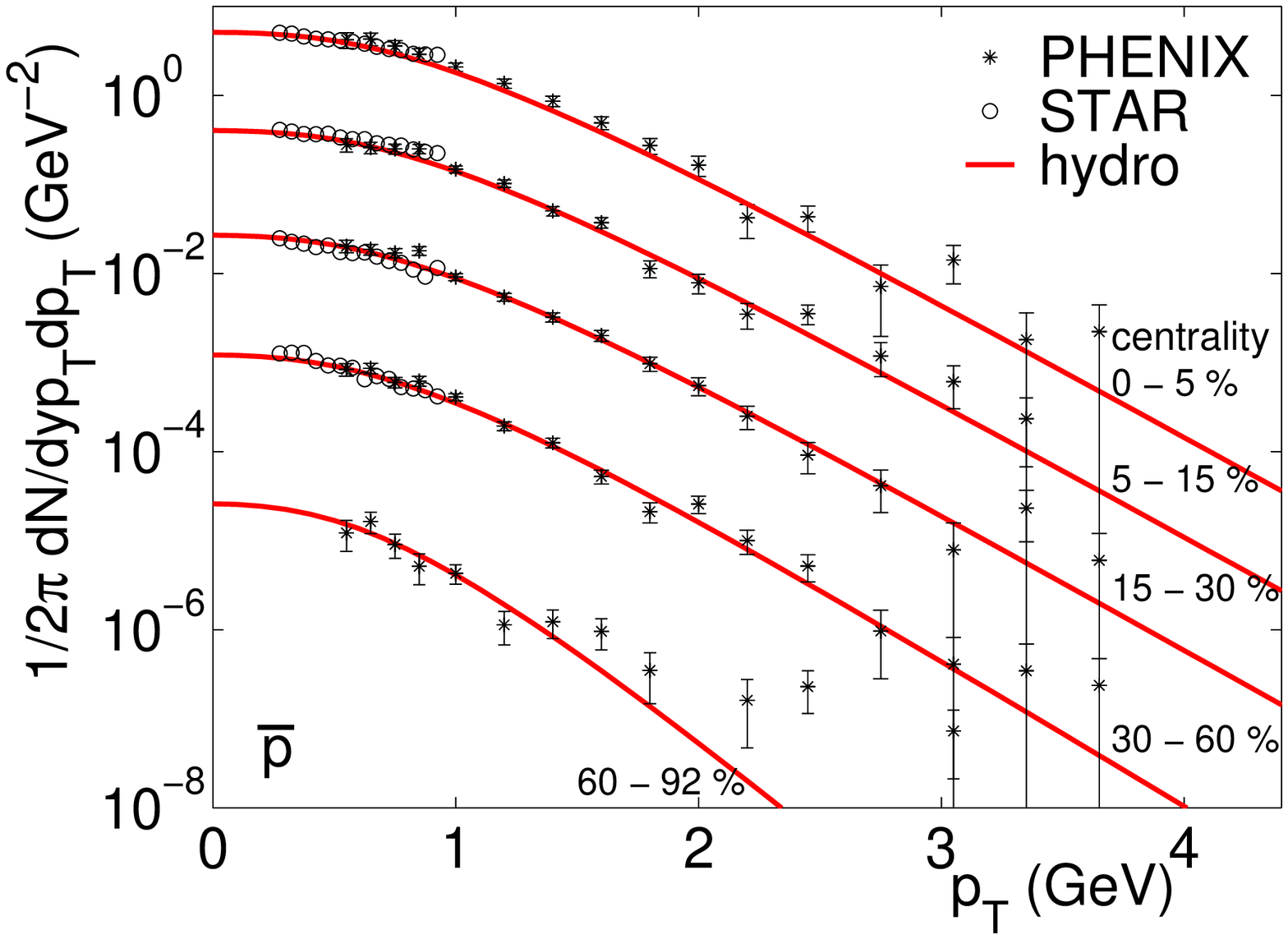,width=60mm,height=50mm}
\hspace*{4mm}\epsfig{file=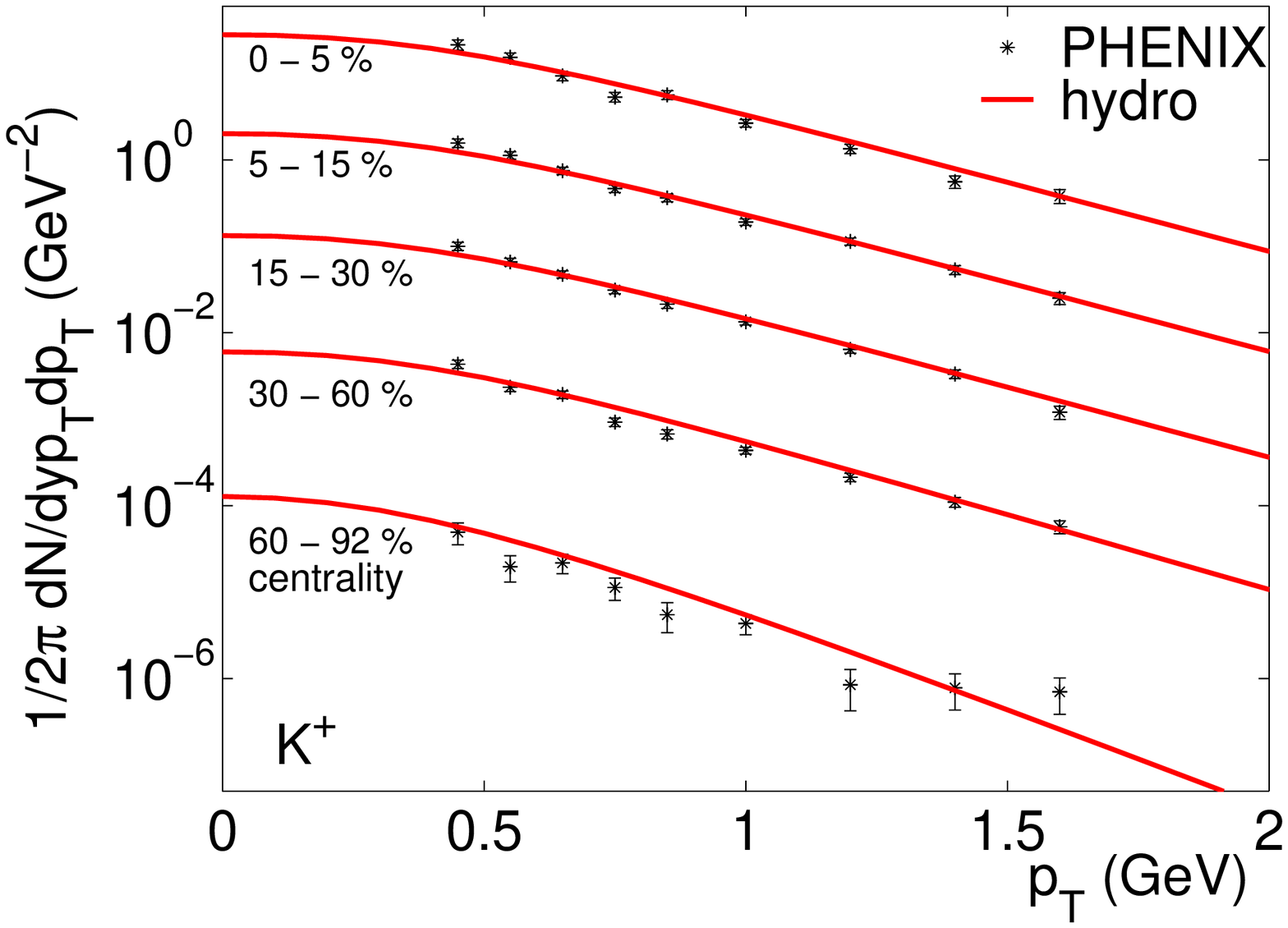,width=60mm,height=50mm}
\vspace*{-5mm}
\caption{\label{F1} 
\small
Charged pion, antiproton and positive kaon spectra from central (upper left
panel) and semi-central to peripheral (other three panels) Au+Au 
collisions at $\sqrt{s}\eq130\,A$\,GeV. The data were taken by the PHENIX 
\cite{PHENIX_spec} and STAR \cite{STAR_spec} collaborations (the STAR data
have slightly different centralities than the indicated values from PHENIX). 
The curves show hydrodynamical calculations (see text).
\vspace*{-6mm}}
\end{figure}

Fig.~\ref{F1} shows the (absolutely normalized) single particle 
$p_\perp$-spectra for charged pions, antiprotons and kaons measured in 
Au+Au collisions at RHIC together with the hydrodynamical results. The 
latter were normalized in central collisions as described above, but 
their centrality dependence and shapes are then completely fixed by the 
model. The agreement with the data is impressive; for antiprotons the data 
go out to $p_\perp{\,\leq\,}3$\,GeV/$c$, and the hydrodynamic model still 
works within errors! Only for very peripheral collisions (impact parameter 
$b{\,>\,}10$\,fm) the data show a significant excess of high-$p_\perp$ 
particles at $p_\perp{\,>\,}1.5$\,GeV/$c$. Teaney {\em et al.} 
\cite{Teaney:2001cw} showed that this excellent agreement {\em requires} 
a phase transition (soft region) in the equation of state; without the 
transition, the agreement is lost, especially when the constraints from 
SPS data and from the elliptic flow measurements below are taken into 
account.

In the hydrodynamic picture the fact that antiprotons become more abundant 
than pions for $p_\perp>2$\,GeV/$c$ (upper left panel of Fig.~\ref{F1}) is 
not surprising at all: it is a simple consequence of the strong radial flow 
at RHIC. For a hydrodynamically expanding thermalized fireball, at 
relativistic transverse momenta $p_\perp{\,\gg\,}m_0$ all hadron spectra 
have the same shape \cite{Lee:1990sk}, and at fixed $m_\perp{\,\gg\,}m_0$ 
their relative normalization is given by $(g_i\lambda_i)/(g_j\lambda_j)$ 
(where $g_{i,j}$ is the spin-isospin degeneracy factor and 
$\lambda_{i,j}=e^{\mu_{i,j}/T}$ is the fugacity of hadron species $i,j$). 
At RHIC the baryon chemical potential at chemical freeze-out is small, 
$\mu_B/T_{\rm chem}{\,\approx\,}0.26$ \cite{Braun-Munzinger:2001ip}, and 
$\mu_\pi\eq0$; the $\bar p/\pi^-$ ratio at fixed and sufficiently large 
$m_\perp$ is thus {\em predicted} to be larger than 1: 
$(\bar p/\pi^-)_{m_\perp}\eq2\exp[-(\mu_B{+}\mu_\pi)/T_{\rm
 chem}]{\,\approx\,}1.5$ (where the factor 2 arises from the spin 
degeneracy of the $\bar p$).

\begin{figure}[htb]
\epsfig{file=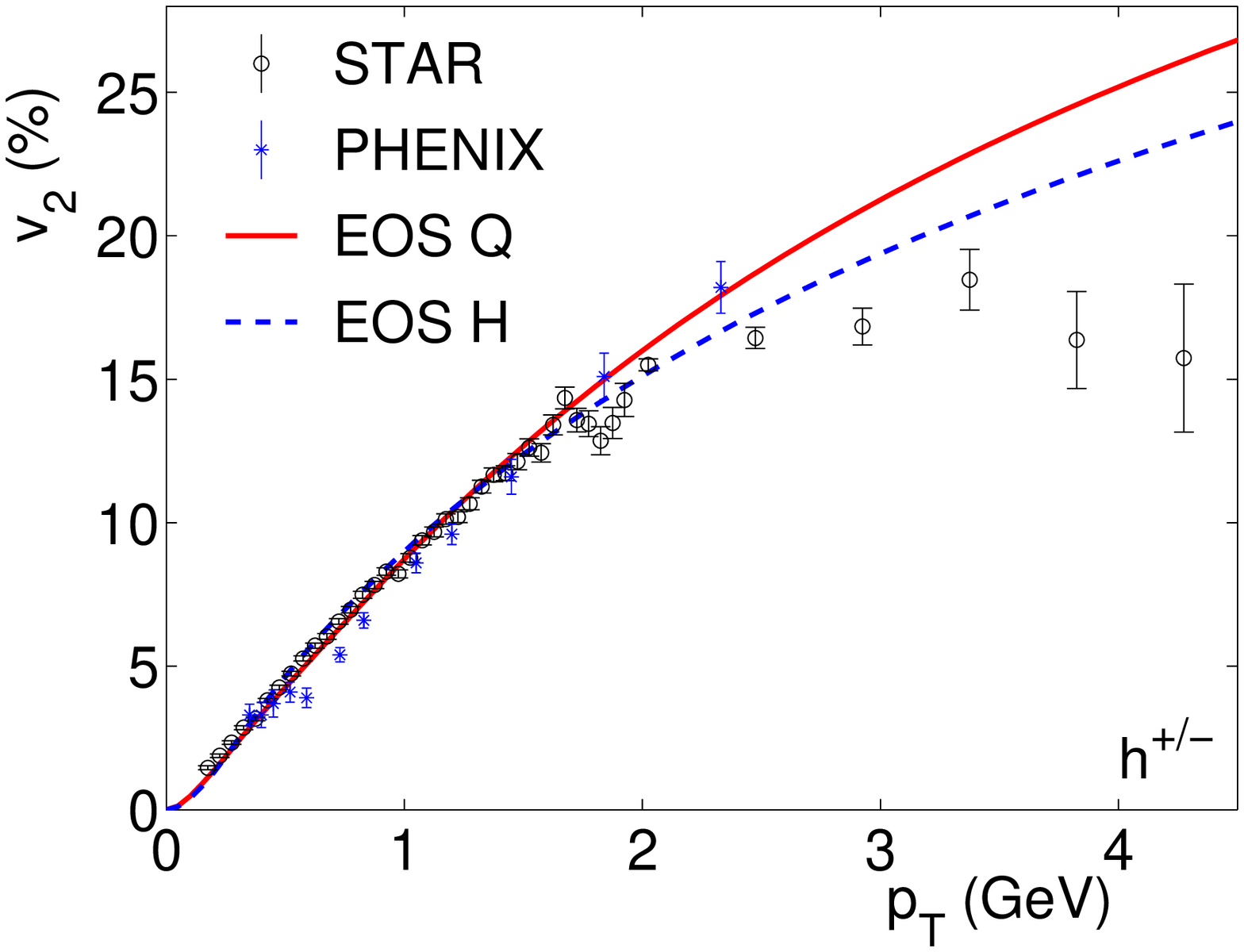,width=62mm,height=50mm}\hspace*{4.5mm}
\epsfig{file=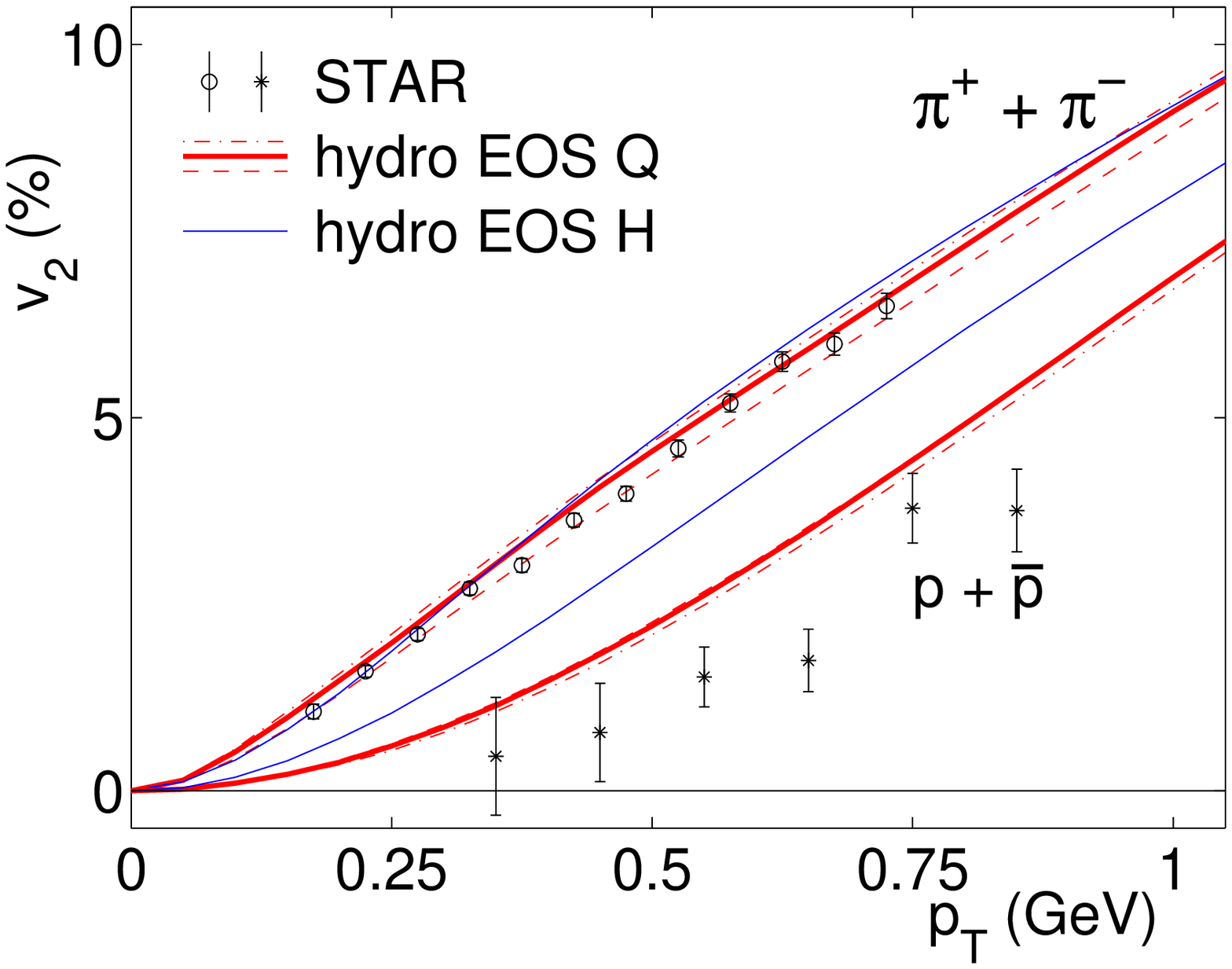,width=58mm,height=50mm}\hfill{\phantom{n}}
\vspace*{-5mm}
\caption{\label{F2} 
\small
The elliptic flow coefficient $\v_2(p_\perp)$ for all charged particles 
(left) and for identified pions and protons (right) from 130\,$A$\,GeV 
minimum bias Au+Au collisions 
\cite{Ackermann:2001tr,Lacey:2001va,Snellings:2001nf,Adler:2001nb}.
The curves are hydrodynamic calculations corresponding to equations of 
state with (Q) and without (H) a phase transition and (in the right panel) 
three different freeze-out temperatures ($T_{\rm f}\eq128$\,MeV (dash-dotted),
130\,MeV (solid) and 134\,MeV (dashed)).
\vspace*{-4mm}
}
\end{figure}

Figure~\ref{F2} compares the differential elliptic flow $\v_2(p_\perp)$ 
from minimum bias Au+Au collisions at RHIC 
\cite{Ackermann:2001tr,Snellings:2001nf,Adler:2001nb}
with hydrodynamic calculations. For transverse momenta 
$p_\perp{\,<\,}2$\,GeV/$c$ the data are seen to exhaust the 
upper limit for $\v_2$ obtained from the hydrodynamic calculations. 
Only for hadrons with $p_\perp>2$\,GeV/$c$ (i.e. fewer than 1\% of all) 
$\v_2$ stays below the hydrodynamic upper limit, indicating incomplete 
thermalization of high-$p_\perp$ particles. The data also show the
hydrodynamically predicted mass-dependence of $\v_2$ 
\cite{Huovinen:2001cy} (right panel). Similar conclusions can be 
drawn from a plot of the $p_\perp$-integrated elliptic flow vs. centrality 
\cite{Kolb:2001fh,KHHET}: only for large impact parameters $b{\,>\,}7$\,fm 
the measured $\v_2$ remains significantly below the hydrodynamic prediction, 
presumably indicating a lack of early thermalization when the initial 
overlap region becomes too small. 

The excellent agreement with hydrodynamics becomes even more 
impressive after you begin to realize how easily it is destroyed: 
As stressed in Sec.~\ref{sec1}, it requires the build-up of momentum
anisotropies during the very early collision stages when the spatial
anisotropy of the reaction zone is still appreciable, causing significant
anisotropies of the pressure gradients. A delay in thermalization by more 
than about 1\,fm/$c$ (2\,fm/$c$) dilutes the spatial anisotropy and the 
hydrodynamically predicted elliptic flow coefficient by 10\% (25\%) 
\cite{Kolb:2000sd} which is more than is allowed by the data. Parton cascade
simulations with standard HIJING input generate almost no elliptic flow
and require an artificial increase of the opacity of the partonic matter
by a factor 80 to reproduce the RHIC data \cite{Molnar:2001ux}. Hadronic
cascades of the RQMD and URQMD type (in which the high-density initial 
state is parametrized by non-interacting, pressureless QCD strings) 
predict \cite{Bleicher:2000sx} too little elliptic flow and a decrease of 
$\v_2$ from SPS to RHIC, contrary to the data. One can get close to the
data by forcing the strings to ``melt'' into partons which then rescatter
with large cross sections \cite{Lin:2001zk}. What causes these large cross 
sections is still a mystery.

The elliptic flow is self-quenching \cite{Sorge:1997pc}: it makes the 
reaction zone grow faster along its initially short direction and thus
eventually eliminates its own cause. As the spatial deformation of the 
fireball goes to zero, the elliptic flow saturates \cite{Kolb:2000sd}.
The saturation time scale times $c$ is of the order of the transverse 
size of the initial overlap region (at lower energies it is a bit longer, 
see Figs.~7,\,9 in \cite{Kolb:2000sd}). At RHIC energies and above, the
time it takes the collision zone to dilute from the high initial energy 
density to the critical value for hadronization is equal to or longer 
than this saturation time: most or all of the elliptic flow is generated 
before any hadrons even appear! It thus seems that the only possible 
conclusion from the successful hydrodynamic description of the observed 
radial and elliptic flow patterns is that the thermal pressure driving 
the elliptic flow is partonic pressure, and that the early stage of the 
collision must have been a thermalized quark-gluon plasma. 

\section{The RHIC HBT puzzle}
\label{sec4}

Hydrodynamics not only predicts the momenta of the emitted hadrons, but
also the spatial structure of the hadron emitting source at freeze-out.
Bose-Einstein (a.k.a. Hanbury Brown-Twiss (HBT)) two-particle intensity 
interferometry allows to access the r.m.s. widths of the space-time 
distribution of hadrons with a given momentum $p$ \cite{Heinz:1999rw}.
One of the interesting questions one can try to address with this tool
is whether at RHIC the reaction zone flips the sign of its spatial 
deformation between initial impact and final freeze-out, as expected 
at very high collision energy \cite{Kolb:2000sd}. The answer to this 
question turns out to be non-trivial, on two different levels: first, 
hydrodynamics, at least with the presently implemented initial conditions 
and freeze-out algorithm, fails to reproduce even for central Au+Au 
collisions the measured HBT radii extracted from two-pion correlations 
\cite{Adler:2001zd,Johnson:2001zi}. We'll show how and explain why. 
Second, for expanding systems the HBT radii don't measure the entire 
freeze-out region, but only the effective emission regions (``regions 
of homogeneity'') for particles of given momentum \cite{Heinz:1999rw}. 
For non-central collisions, due to the anisotropic transverse flow 
these can, at least in principle, have a different spatial deformation 
than the entire (momentum-integrated) freeze-out region, giving rise 
to a different behaviour of the HBT radii observed at different angles 
relative to the reaction plane than perhaps naively expected 
\cite{Heinz:2001xi}.

\begin{figure}[htb]
\epsfig{file=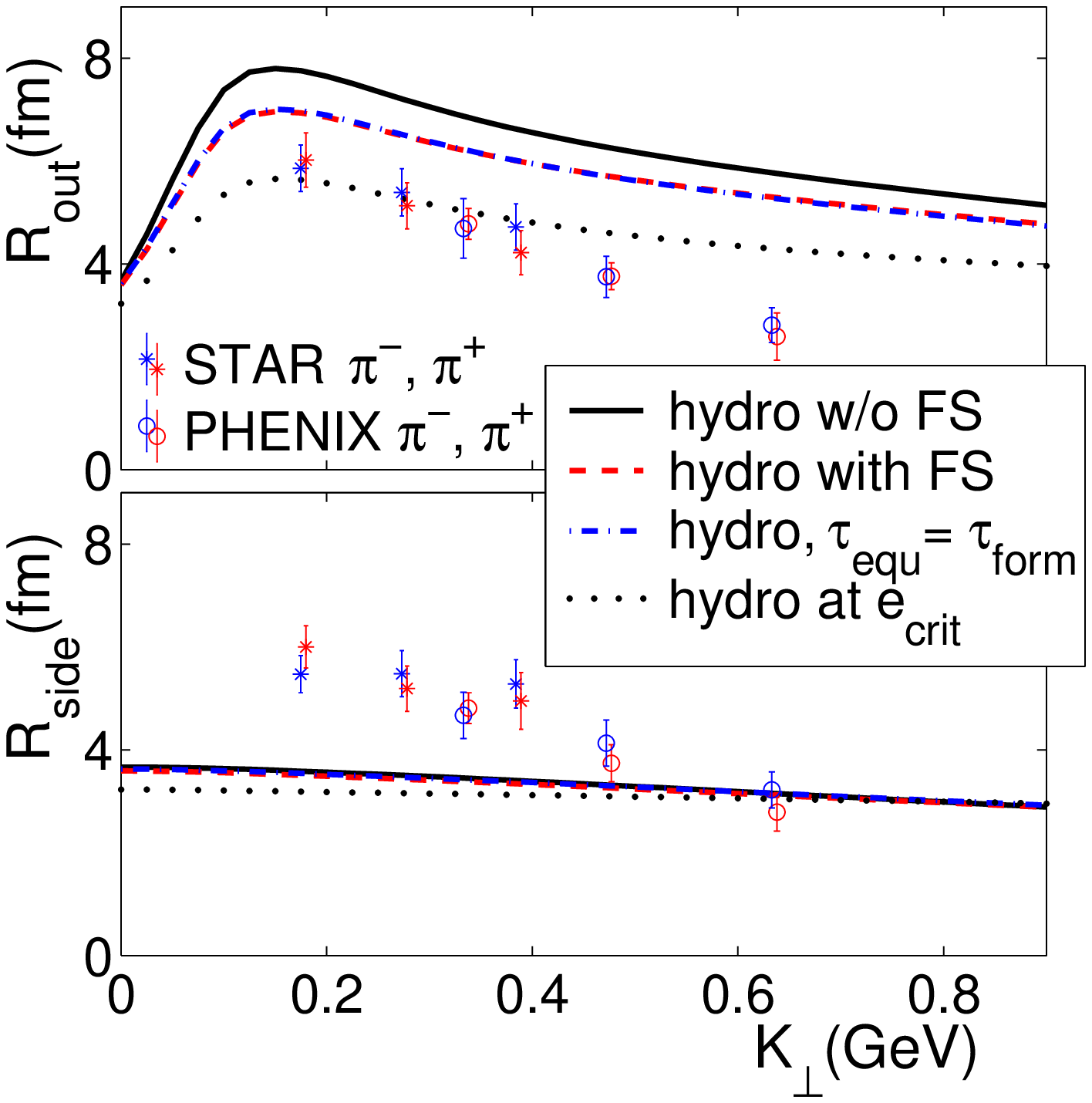,width=60mm,height=60mm}\hspace*{1.5mm}
\epsfig{file=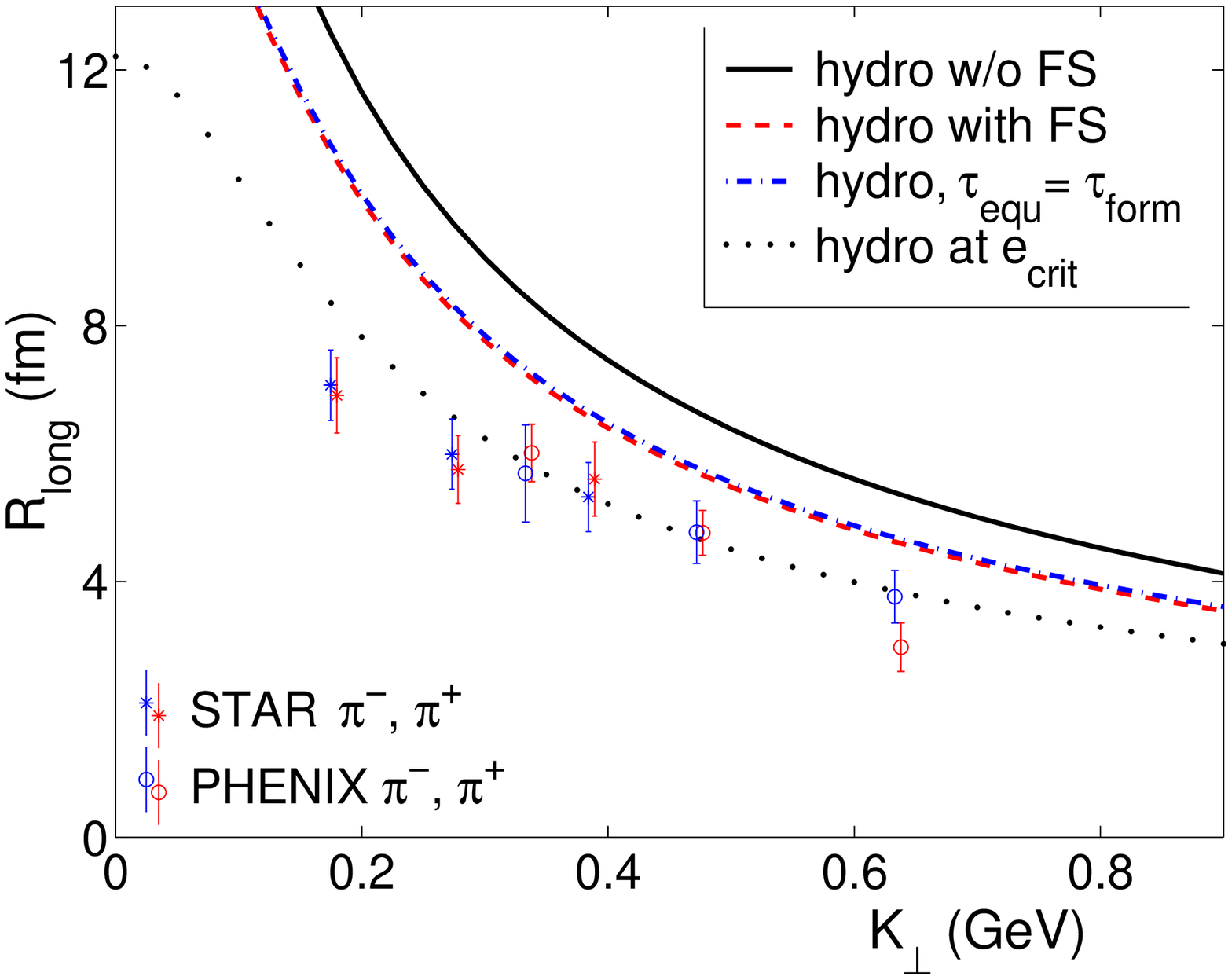,width=62mm,height=60mm}\hfill{\phantom{n}}
\vspace*{-5mm}
\caption{\label{F3} 
\small
HBT radii from a hydrodynamic source compared to RHIC data 
\cite{Adler:2001zd,Adcox:2002uc}. The solid lines show hydrodynamic
results with standard initialization and freeze-out (see text). The 
dotted lines assume freeze-out directly after hadronization at 
$e_{\rm dec}\eq{e}_{\rm crit}$. The other lines correspond to 
modified initial conditions as described in the text. 
\vspace*{-5mm}
}
\end{figure}

Fig.~\ref{F3} shows the HBT radii $R_{\rm side}$, $R_{\rm out}$, and 
$R_{\rm long}$ for central Au+Au collisions at $\sqrt{s}\eq130\,A$\,GeV
\cite{Adler:2001zd,Adcox:2002uc} compared with hydrodynamic results. The 
theoretical results only account for directly emitted pions; resonance
decay pions would add an exponential tail to the emission time distribution 
and could cause a slight increase in the emission duration, without affecting 
much the spatial width of the emission region \cite{Wiedemann:1996ig}.
The solid lines correspond to initial conditions and freeze-out parameters
as used for the successful description of the momentum distributions in 
the previous section; clearly, $R_{\rm side}$ comes out too small whereas 
$R_{\rm out}$ and $R_{\rm long}$ are both too large in the model. The 
problem of $R_{\rm side}$ from hydrodynamics being too small is well-known 
from the SPS \cite{Schlei:1996mc}; presumably it is mostly due to the sharp 
Cooper-Frye freeze-out and seems to be at least partially resolved if 
the freeze-out kinetics is handled microscopically within a hadronic 
cascade \cite{Soff:2001eh}. The latter gives a more ``fuzzy'' spatial 
freeze-out distribution with larger r.m.s. width in the sideward 
direction; it is not so clear that it also resolves the problem with the
$K_\perp$-dependence of $R_{\rm side}$ which is much stronger in the data
than in the model.  
 
On the other hand, ``fuzzy'' freeze-out \cite{Soff:2001eh} and the 
inclusion of resonance decay contributions \cite{Wiedemann:1996ig} 
only exacerbate the problems with $R_{\rm long}$ and $R_{\rm out}$. 
In high energy heavy-ion collisions 
the longitudinal HBT radius is controlled by the expansion dynamics via 
the longitudinal velocity gradient at freeze-out \cite{Heinz:1999rw}. For 
a boost invariant longitudinal flow profile this gradient decreases with 
time as $1/\tau$, leading to rather weak gradients (and correspondingly 
large values for $R_{\rm long}$) at the typical hydrodynamic freeze-out 
time of ${\sim\,}15$\,fm/$c$. To reduce $R_{\rm long}$, either freeze-out 
must happen significantly earlier, or the expansion of the source must 
strongly violate longitudinal boost-invariance even close to mid-rapidity, 
somehow allowing for significantly stronger longitudinal flow 
velocity gradients at late times than predicted by the Bjorken profile
$v_L{\eq}z/t$.  

Noting that decoupling is driven by the strong radial flow developing
during the expansion stage \cite{Schnedermann:gc}, we have tried to force 
the system to decouple earlier by initiating the transverse flow even 
before $\tau_{\rm eq}$. Even if the system is not yet locally thermalized,
it will start to develop some transverse collective dynamics, albeit perhaps
not as quickly as in a hydrodynamic approach. We have tested two extreme 
assumptions about the transverse expansion prior to thermalization \cite{KTH}: 
in one simulation, shown as the dot-dashed curve in Fig.~\ref{F3}, we started 
the hydrodynamic evolution directly at the parton formation time (for which 
we took the somewhat arbitrary value $\tau_{\rm form}\eq0.2$\,fm/$c$). In 
another limit (dashed lines in Fig.~\ref{F3}), we allowed the partons to
stream freely from time $\tau_{\rm form}$ to $\tau_{\rm eq}$ and matched 
at $\tau_{\rm eq}$ the first row of the energy momentum tensor to an ideal 
fluid form, thereby extracting an initial transverse flow profile at 
$\tau_{\rm eq}$. In both cases the resulting transverse flow ``seed'' at 
$\tau_{\rm eq}\eq0.6$\,fm/$c$ caused the system to expand more rapidly and 
farther out into the transverse direction, freezing out 10-20\% earlier.
Fig.~\ref{F3} shows that this helps with both $R_{\rm long}$ and 
$R_{\rm out}$, but not as much as required by the data. 

To get close to the data, we would need to postulate freeze-out directly 
at the hadroni\-zation point (dotted lines in Fig.~\ref{F3}). This radical 
and entirely unmotivated assumption almost removes the discrepancy with 
$R_{\rm long}$ and with the average magnitude of $R_{\rm out}$, but it is 
still unable to reproduce the measured strong $K_\perp$-depenence of 
$R_{\rm out}$ and $R_{\rm side}$, and it still yields much too small
values for $R_{\rm side}$ at low $K_\perp$. Correspondingly, the ratio
$R_{\rm out}/R_{\rm side}$ (which is $\approx 1$ in the data) is still
significantly overpredicted by the model (although much less so than by 
earlier hydrodynamic predictions \cite{Rischke:1996em} when constraints 
from RHIC momentum spectra were not yet available).

The problem with the small $R_{\rm side}$ cannot be fixed by pre-equilibrium 
transverse flow, either: even though the system then expands to larger 
values of $r$, $R_{\rm side}$ remains essentially unchanged (dashed
and dash-dotted lines in Fig.~\ref{F3}). The homogeneity region simply
moves farther out without increasing in size. We don't see a way to move
closer to the data by further modifying the initial conditions, and even
the alterations we made to obtain Fig.~\ref{F3} may turn out to be
excluded by the singles spectra and the elliptic flow data (which we 
haven't tested yet). We therefore believe that a resolution to the HBT
puzzle must lie in the handling of the freeze-out process (although we 
don't yet know how). 

\begin{figure}[htb]
\vspace*{-2mm}
\epsfig{file=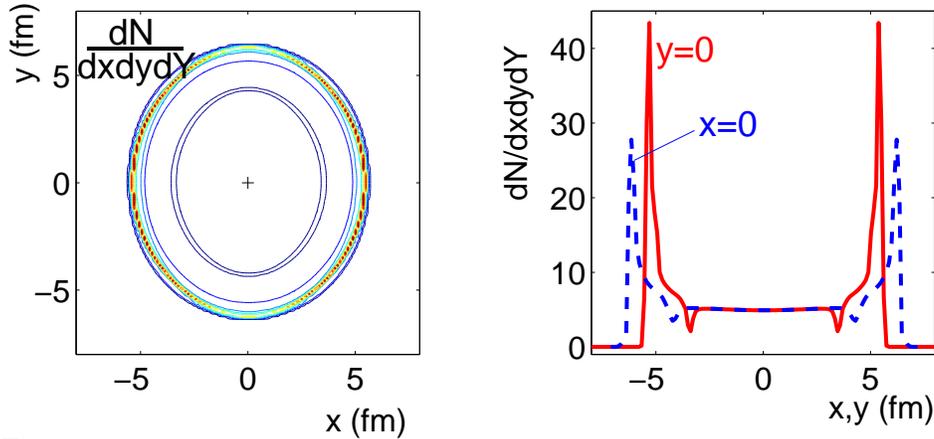,width=125mm}
\vspace*{-5mm}
\caption{\label{F4} 
\small
The $K_\perp$-integrated emission function from hydrodynamics, for 
directly emitted pions with rapidity $Y=0$ from semicentral Au+Au 
collisions ($b\eq7$\,fm). The emission function is integrated over time 
and longitudinal coordinate $z$. The left panel shows contours of 
constant particle density in the transverse plane, the right panel 
presents cuts through this distribution along the $x$ and $y$ axes,
showing that particle emission is strongly concentrated at the surface,
especially for emission into the reaction plane ($y\eq0$). 
\vspace*{-5mm}
}
\end{figure}

\begin{figure}[ht]
\begin{center}
\vspace*{-3mm}
\epsfig{file=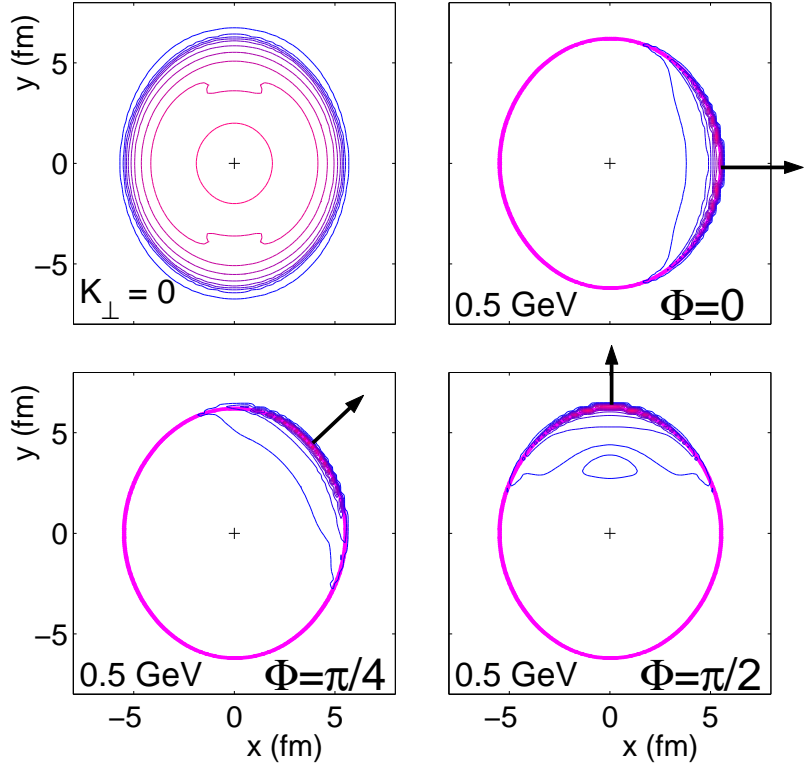,width=107mm}
\end{center}
\vspace*{-5mm}
\caption{\label{F5} 
\small
Time- and $z$-integrated emission function for $b\eq7$\,fm Au+Au 
collisions at RHIC for directly emitted pions with rapidity $Y\eq0$ 
and fixed transverse momentum $K_\perp$. Shown are contours of constant
particle density at freeze-out in the transverse plane, for $K_\perp\eq0$
in the upper left panel, and for $K_\perp\eq0.5$\,GeV and three azimuthal
emission angles ($0^\circ$, $45^\circ$, and $90^\circ$ relative to the 
reaction plane) in the remaining three panels.
\vspace*{-5mm}
}
\end{figure}

Let us further analyze the relationship between $R_{\rm out}$ and 
$R_{\rm side}$. The formula \cite{Heinz:1999rw}
 \begin{eqnarray}
 \label{Ros}
  R_{\rm out}^2-R_{\rm side}^2 = 
  \langle(x_{\rm out}{-}\bar x_{\rm out})^2 
       - (x_{\rm side}{-}\bar x_{\rm side})^2\rangle + 
  \beta_\perp^2\langle(t{-}\bar t)^2\rangle -
  2\beta_\perp\langle(x_{\rm out}{-}\bar x_{\rm out})(t{-}\bar t)\rangle
 \end{eqnarray}
shows that their difference is controlled by three contributions which
can partially compensate each other. There is always a positive
contribution from the emission duration $(\delta t)^2{\,\equiv\,}\langle
 (t{-}\bar t)^2\rangle$; for systems with longitudinal boost-invariance
freeze-out happens at constant longitudinal proper time $\tau\eq\tau_{\rm f}$,
and late freeze-out, which leads to a large longitudinal homogeneity region 
$R_{\rm long}$, thus implies contributions from a wide region along 
this $\tau\eq\tau_{\rm f}$ hyperbola, leading to a large variance 
$\delta t$. To reduce $\delta t$ we should therefore again freeze out 
earlier. In hydrodynamics with Cooper-Frye freeze-out \cite{Cooper:1974mv}, 
the term 
$-2\beta_\perp\langle(x_{\rm out}{-}\bar x_{\rm out})(t{-}\bar t)\rangle$ 
is also positive, since freeze-out happens from the outside inward, causing
a negative $x_{\rm out}{-}t$ correlation. These two positive contributions to 
$R_{\rm out}^2{-}R_{\rm side}^2$ are partially cancelled by a negative 
geometric contribution $(\delta x_{\rm out})^2{-}(\delta x_{\rm side})^2$: 
according to Figures~\ref{F4} and \ref{F5} hydrodynamics predicts that pion 
emission is strongly surface dominated, especially for pions with non-zero 
transverse momentum $K_\perp$, whose emission regions are tightly squeezed 
towards the edge of the fireball where the radial flow is strongest. This 
results in an ``opaque source'', characterized by a smaller outward than 
sideward variance $(\delta x_{\rm out})^2<(\delta x_{\rm side})^2$ 
\cite{Tomasik:1998qt}. The opacity of the hydrodynamic source is stronger 
at RHIC than SPS energies, but not strong enough to compensate for the 
two positive contributions in Eq.~(\ref{Ros}); this is the reason why we 
fail to reproduce the measured relation $R_{\rm out}\approx R_{\rm side}$. 
Of course, things would be easier if the $x_{\rm out}{-}t$ correlation 
were positive (i.e. pions at larger radial distances were emitted later), 
but we don't see a good reason why this should be so.

\begin{figure}[ht]
\vspace*{-3mm}
\epsfig{file=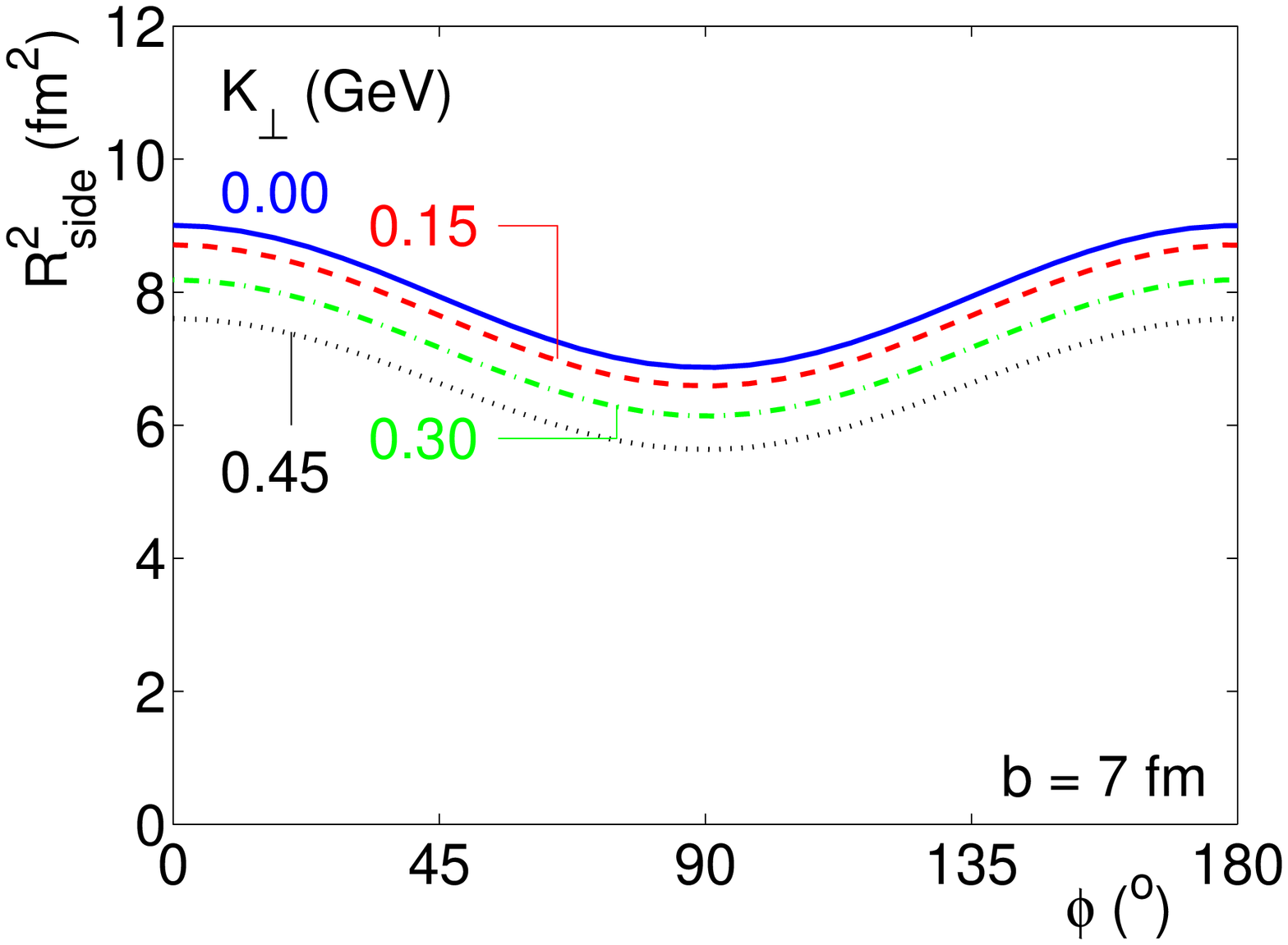,width=62mm,height=50mm}\hspace*{1mm}
\epsfig{file=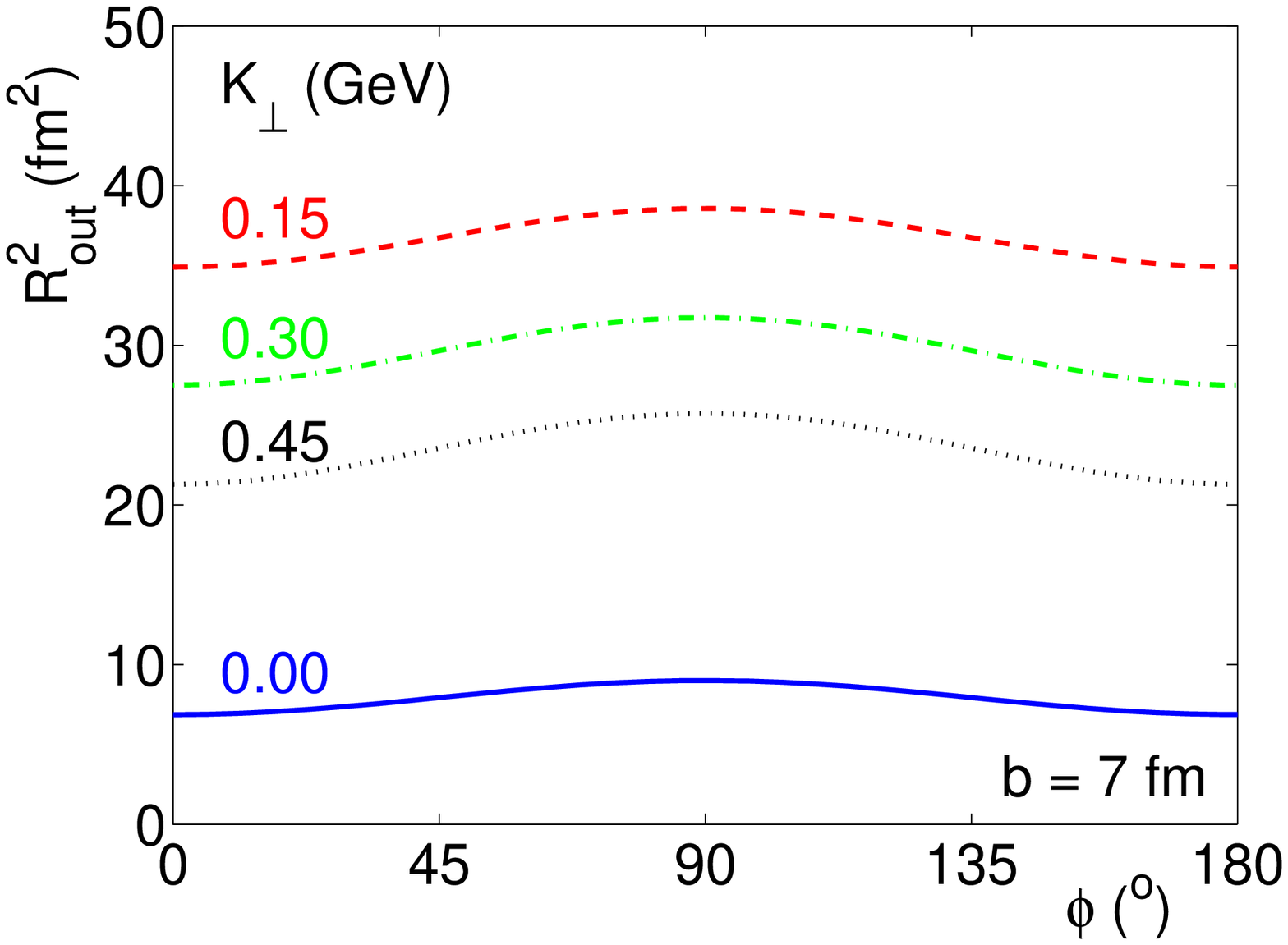,width=61mm,height=50mm}\hfill{\phantom{n}}
\epsfig{file=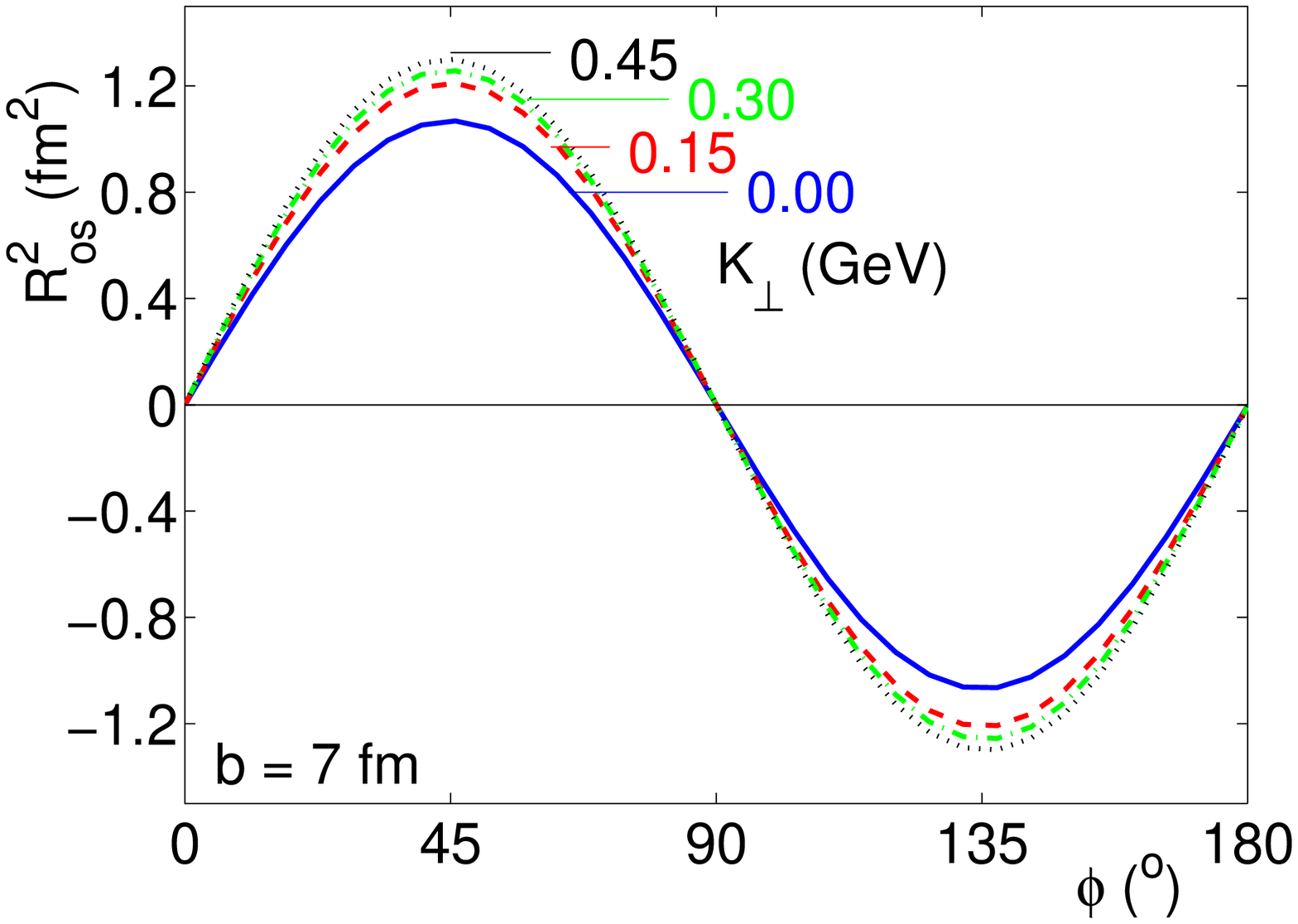,width=62mm,height=50mm}
\epsfig{file=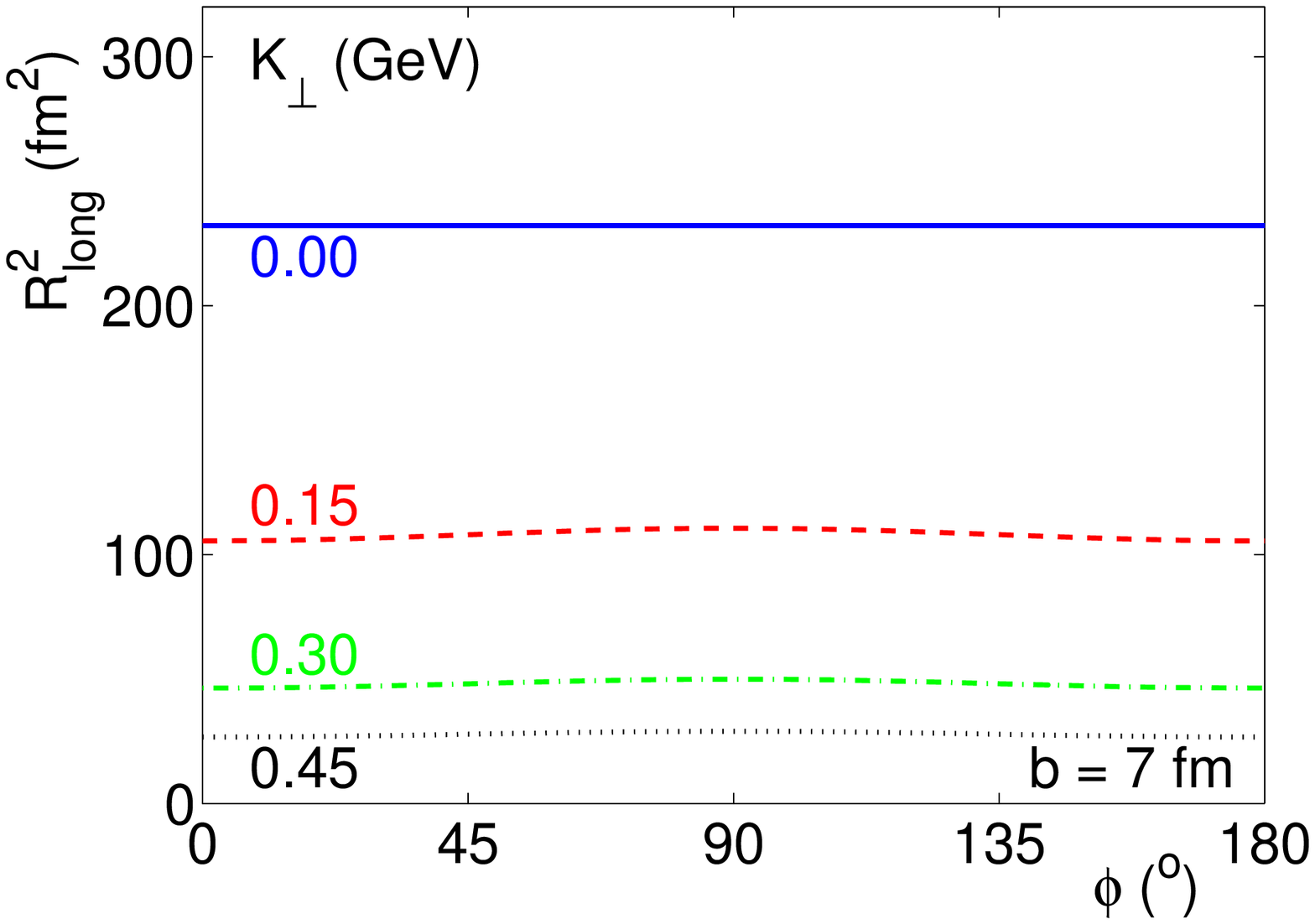,width=62mm,height=50mm}
\vspace*{-5mm}
\caption{\label{F6} 
\small
Azimuthal dependence of the HBT radii from hydrodynamic simultations
of RHIC~1 Au+Au collisions at $b\eq7$\,fm, for pion pairs with rapity 
$Y\eq0$ and varius values of $K_\perp$ as indicated in the diagrams.
The oscillations are about 25\% stronger if freeze-out is implemented
directly at hadronization (P. Kolb, PhD thesis).
\vspace*{-5mm}
}
\end{figure}

Let us close with a quick preview of results for the HBT radii for 
non-central collisions, in particular their dependence on the angle 
$\Phi$ of the transverse pair momentum $\vec K_\perp$ relative to the 
reaction plane. Fig.~\ref{F6} shows our hydrodynamic results for 
$R_{\rm side}^2$, $R_{\rm out}^2$, $R_{\rm os}^2$, and $R_{\rm long}^2$ 
for $b\eq7$\,fm Au+Au collisions at $\sqrt{s}\eq130\,A$\,GeV, plotted as 
functions of $\Phi$ for a number of different values of 
$K_\perp{\,=\,}|\vec K_\perp|$. (Note that ``out'' and ``side'' denote 
the directions parallel and perpendicular to $\vec K_\perp$ in the 
transverse plane \cite{Heinz:1999rw}.) While $R_{\rm long}^2$ is almost 
independent of $\Phi$, the three other radius parameters show marked 
azimuthal dependences of the generic form (with all coefficients being 
positive)
 \begin{eqnarray}
   &&R_{\rm side}^2(\Phi) = R_{{\rm s},0}^2 + R_{{\rm s},2}^2 \cos(2\Phi),
   \qquad
     R_{\rm out}^2(\Phi) = R_{{\rm o},0}^2 - R_{{\rm o},2}^2 \cos(2\Phi), 
 \nonumber\\
   &&R_{\rm os}^2(\Phi) = R_{{\rm os},2}^2 \sin(2\Phi).
 \end{eqnarray}
Although the magnitudes of the coefficients $R_\alpha^2$ in Fig.~\ref{F6}
are quite different (and presumably not too trustworthy, given the 
disagreement with the data for central collisions in Fig.~\ref{F3}), 
it is surprising that the signs and phases of the oscillations are 
identical to those calculated and measured at the AGS \cite{Lisa:2000ip}! 
At the AGS radial flow effects are thought to be sufficiently weak that 
the oscillations can be interpreted purely geometrically \cite{Lisa:2000ip}, 
reflecting a spatially deformed source which is elongated perpendicular to 
the reaction plane (as is the case for the initial overlap region). 
Fig.~\ref{F5} shows that at RHIC the reason for these oscillations is more
subtle: except for small $K_\perp$, where pions are emitted from the entire
fireball, the strong radial flow squeezes the emission regions towards the 
edge of the fireball so that they no longer reflect directly the spatial 
deformation of the momentum-integrated emission function shown in 
Fig.~\ref{F4}. However, even though the measured HBT radii have little to 
do with the total widths of the fireball in $x$ and $y$ directions, the 
out-of-plane deformation of the latter imprints itself on the shape of 
the effective emission regions through the ``opacity effect'', i.e. the 
fact that the emission is strongly surface-peaked, leading to azimuthal 
oscillations which are in phase with those expected from naive geometrical 
considerations of the entire source. 

Even though the absolute magnitudes of the radii shown in  Fig.~\ref{F6}
are affected by similar problems as those in Fig.~\ref{F3}, the phases
of the azimuthal oscillations agree with preliminary STAR data
\cite{Retiere:2001ed}, and also their amplitudes are at least in the right 
ballpark. This gives rise to some optimism that the HBT puzzle is limited 
in scope and can be resolved by evolutionary rather than revolutionary 
methods.

\section*{Acknowledgement}
This work was supported by the U.S. Department of Energy under contract 
DE-FG02-01ER41190.
 

\vfill\eject

\begin{thebibliography}{99}

\bibitem{O92} 
  J.-Y.~Ollitrault, {\it Phys.~Rev.}~{\bf D46} (1992) 229.

\bibitem{VZ96}
  S.A.~Voloshin and Y.~Zhang, {\it Z.~Phys.} {\bf C70} (1996) 665.

\bibitem{Zhang:1999rs}
B.~Zhang, M.~Gyulassy, and C.M.~Ko,
{\it Phys.\ Lett.} {\bf B455} (1999) 45.

\bibitem{Molnar:2001ux}
D.~Moln\'ar and M.~Gyulassy,
{\it Nucl.~Phys.} {\bf A697} (2002) 495.

\bibitem{Kolb:2000sd}
P.F.~Kolb, J.~Sollfrank and U.~Heinz,
{\it Phys.\ Rev.} {\bf C62} (2000) 054909.

\bibitem{Sorge:1997pc}
H.~Sorge,
{\it Phys.\ Rev.\ Lett.} {\bf 78} (1997) 2309; 
{\em ibid.} {\bf 82} (1999) 2048.

\bibitem{Voloshin:2000gs}
S.A.~Voloshin and A.M.~Poskanzer,
{\it Phys.\ Lett.} {\bf B474} (2000) 27.

\bibitem{Ackermann:2001tr}
K.H.~Ackermann {\it et al.}  [STAR Collaboration],
{\it Phys.\ Rev.\ Lett.} {\bf 86} (2001) 402.

\bibitem{Lacey:2001va}
R.A.~Lacey {\it et al.} [PHENIX Collaboration], 
{\it Nucl. Phys.} {\bf A698} (2002) 559c.

\bibitem{Poskanzer:2001cx}
For a recent compilation of elliptic flow data see 
A.M.~Poskanzer,
nucl-ex/0110013.

\bibitem{Kolb:1999it}
P.F.~Kolb, J.~Sollfrank, and U.~Heinz,
{\it Phys.\ Lett.} {\bf B459} (1999) 667.

\bibitem{Teaney:2001cw}
D.~Teaney, J.~Lauret, and E.V.~Shuryak,
{\it Phys.\ Rev.\ Lett.}\ {\bf 86} (2001) 4783;
and nucl-th/0110037.

\bibitem{Kolb:2001fh}
P.F.~Kolb {\it et al.}, 
{\it Phys.\ Lett.} {\bf B500} (2001) 232.

\bibitem{Huovinen:2001cy}
P.~Huovinen {\it et al.},
{\it Phys. Lett.} {\bf B503} (2001) 58.

\bibitem{Adler:2001zd}
C.~Adler {\it et al.} [STAR Collaboration],
{\it Phys.\ Rev.\ Lett.} {\bf 87} (2001) 082301.

\bibitem{Johnson:2001zi}
S.C.~Johnson [PHENIX Collaboration], 
{\it Nucl. Phys.} {\bf A698} (2002) 603c.

\bibitem{Cooper:1974mv}
F.~Cooper and G.~Frye,
{\it Phys.\ Rev.} {\bf D10} (1974) 186.

\bibitem{Bass:2000ib}
S.A.~Bass and A.~Dumitru,
{\it Phys.\ Rev.} {\bf C61} (2000) 064909.

\bibitem{Heinz:1999kb}
U.~Heinz,
{\it Nucl.\ Phys.} {\bf A661} (1999) 140c.

\bibitem{Bjorken:1983qr}
J.D.~Bjorken,
{\it Phys. Rev.} {\bf D27} (1983) 140.

\bibitem{KHHET}
  P.F.~Kolb, U. Heinz, P. Huovinen, K.J. Eskola, and K. Tuominen, 
  {\it Nucl. Phys.} {\bf A696} (2001) 175.

\bibitem{PHOBOS}
  B.B.~Back {\it et al.} [PHOBOS Collaboration], 
  {\it Phys. Rev. Lett.} {\bf 85} (2000) 3100; 
  {\it Phys. Rev.} {\bf C65} (2002) 031901; 
  and {\it Phys. Rev. Lett.} {\bf 88} (2002) 022302.

\bibitem{PHENIX}
  K.~Adcox {\it et al.} [PHENIX Collaboration], 
  {\it Phys. Rev. Lett.} {\bf 86} (2001) 3500.

\bibitem{PHENIX_spec}
  J.~Velkovska {\it et al.} [PHENIX Collaboration], 
  {\it Nucl. Phys.} {\bf A698} (2002) 507c.

\bibitem{STAR_spec}
  C.~Adler {\it et al.} [STAR Collaboration],
  {\it Phys. Rev. Lett.} {\bf 87} (2001) 262302.

\bibitem{Lee:1990sk}
  K.S.~Lee, U.~Heinz, and E.~Schnedermann,
{\it Z.\ Phys.} {\bf C48} (1990) 525.

\bibitem{Braun-Munzinger:2001ip}
  P.~Braun-Munzinger {\it et al.},
  {\it Phys.\ Lett.} {\bf B518} (2001) 41.

\bibitem{Snellings:2001nf}
  R.J.~Snellings  [STAR Collaboration],
  {\it Nucl. Phys.} {\bf A698} (2002) 193c.

\bibitem{Adler:2001nb}
  C.~Adler {\it et al.} [STAR Collaboration],
  {\it Phys.\ Rev.\ Lett.} {\bf 87} (2001) 182301.

\bibitem{Bleicher:2000sx}
  M.~Bleicher and H.~St\"ocker,
  {\it Phys. Lett.} {\bf B526} (2002) 309.

\bibitem{Lin:2001zk}
  Z.W.~Lin and C.M.~Ko,
  {\it Phys.\ Rev.} {\bf C65} (2002) 034904.

\bibitem{Heinz:1999rw}
U.~Heinz and B.V.~Jacak,
{\it Ann. Rev. Nucl. Part. Sci.} {\bf 49} (1999) 529;
U.A.~Wiedemann and U.~Heinz,
{\it Phys. Rept.} {\bf 319} (1999) 145.

\bibitem{Heinz:2001xi}
U.~Heinz and P.F.~Kolb,
hep-ph/0111075.

\bibitem{Adcox:2002uc}
K.~Adcox {\it et al.}  [PHENIX Collaboration],
nucl-ex/0201008.

\bibitem{Wiedemann:1996ig}
U.A.~Wiedemann and U.~Heinz,
{\it Phys. Rev.} {\bf C56} (1997) 3265.

\bibitem{Schlei:1996mc}
B.R.~Schlei and N.~Xu,
{\it Phys. Rev.} {\bf C54} (1996) 2155;
B.R.~Schlei, D.~Strottman, J.P.~Sullivan, and H.W.~van Hecke,
{\it Eur. Phys. J.} {\bf C10} (1999) 483.

\bibitem{Soff:2001eh}
S.~Soff, S.A.~Bass, and A.~Dumitru,
{\it Phys. Rev. Lett.} {\bf 86} (2001) 3981.

\bibitem{Schnedermann:gc}
E.~Schnedermann and U.~Heinz,
{\it Phys. Rev.} {\bf C50} (1994) 1675.

\bibitem{KTH}
  P.F.~Kolb, M.~Tilley, and U.~Heinz, in preparation. 

\bibitem{Rischke:1996em}
D.H.~Rischke and M.~Gyulassy,
{\it Nucl. Phys.} {\bf A608} (1996) 479.

\bibitem{Tomasik:1998qt}
B.~Tom\'asik and U.~Heinz,
nucl-th/9805016; and 
{\it Acta Phys. Slov.} {\bf 49} (1999) 251 [nucl-th/9901006].

\bibitem{Lisa:2000ip}
M.A.~Lisa, U.~Heinz, and U.A.~Wiedemann,
{\it Phys. Lett.} {\bf B489} (2000) 287;
M.A.~Lisa {\it et al.} [E895 Collaboration],
{\it Phys. Lett.} {\bf B496} (2000) 1.

\bibitem{Retiere:2001ed}
F.~Reti\`ere  [STAR Collaboration],
nucl-ex/0111013.

\end{thebibliography}
\end{document}